\documentclass[prx,a4paper,aps,twocolumn,superscriptaddress,10pt,bibnotes]{revtex4-1}
\usepackage[english]{babel}
\usepackage[utf8]{inputenc}

\setcitestyle{sort&compress,numbers}
\usepackage{amsthm, color, mathtools, amsmath, amssymb, setspace,bbm, tensor, subfigure, mathtools, placeins, graphicx, wrapfig,float, verbatim, xcolor,enumitem}
\usepackage{graphicx}
\usepackage[unicode=true, breaklinks=false, pdfborder={0 0 1}, backref=false, colorlinks=true, linkcolor=blue, citecolor=blue]{hyperref}
\usepackage{cancel,bm}

\let\oldsqrt\sqrt
\def\sqrt{\mathpalette\DHLhksqrt}
\def\DHLhksqrt#1#2{%
\setbox0=\hbox{$#1\oldsqrt{#2\,}$}\dimen0=\ht0
\advance\dimen0-0.2\ht0
\setbox2=\hbox{\vrule height\ht0 depth -\dimen0}%
{\box0\lower0.4pt\box2}}

\newcommand{\tr}{\operatorname{tr}}
\newcommand{\ptr}[1]{\operatorname{tr}_{#1}}

\newcommand{\inp}{\text{\normalfont \texttt{i}}}
\newcommand{\out}{\text{\normalfont \texttt{o}}}

\def\bra#1{\mathinner{\langle{#1}|}}

\def\ket#1{\mathinner{|{#1}\rangle}}

\newcommand*\xbar[1]{%
   \hbox{%
     \vbox{%
       \hrule height 0.5pt 
       \kern0.5ex
       \hbox{%
         \kern-0.2em
         \ensuremath{#1}%
         \kern-0.0em
       }%
     }%
   }%
} 

\def\BraVert{\egroup\,\mid\,\bgroup}

\def\bra#1{\mathinner{\langle{#1}|}}

\def\ket#1{\mathinner{|{#1}\rangle}}

\usepackage{bbm, braket}
\usepackage[mathscr]{euscript}
\allowdisplaybreaks
\newtheorem*{theorem*}{Theorem}

\newtheorem*{corollary*}{Corollary}

\begin{document}

\title{Experimental Demonstration of Instrument-specific Quantum Memory Effects and Non-Markovian Process Recovery for Common-Cause Processes}

\author{Yu Guo}
\thanks{These two authors contributed equally.}
\affiliation{CAS Key Laboratory of Quantum Information, University of Science and Technology of China, Hefei 230026, People's Republic of China}
\affiliation{CAS Center for Excellence in Quantum Information and Quantum Physics, University of Science and Technology of China, Hefei 230026, People's Republic of China}

\author{Philip Taranto}
\thanks{These two authors contributed equally}
\email{philip.taranto@oeaw.ac.at} 
\affiliation{Institute for Quantum Optics and Quantum Information - IQOQI Vienna, Austrian Academy of Sciences, Boltzmanngasse 3, 1090 Vienna, Austria}
\affiliation{Atominstitut, Technische Universit{\"a}t Wien, 1020 Vienna, Austria}

\author{Bi-Heng Liu}
\email{bhliu@ustc.edu.cn}
\affiliation{CAS Key Laboratory of Quantum Information, University of Science and Technology of China, Hefei 230026, People's Republic of China}
\affiliation{CAS Center for Excellence in Quantum Information and Quantum Physics, University of Science and Technology of China, Hefei 230026, People's Republic of China}

\author{Xiao-Min Hu}
\affiliation{CAS Key Laboratory of Quantum Information, University of Science and Technology of China, Hefei 230026, People's Republic of China}
\affiliation{CAS Center for Excellence in Quantum Information and Quantum Physics, University of Science and Technology of China, Hefei 230026, People's Republic of China}

\author{Yun-Feng Huang}
\affiliation{CAS Key Laboratory of Quantum Information, University of Science and Technology of China, Hefei 230026, People's Republic of China}
\affiliation{CAS Center for Excellence in Quantum Information and Quantum Physics, University of Science and Technology of China, Hefei 230026, People's Republic of China}

\author{Chuan-Feng Li}
\email{cfli@ustc.edu.cn}
\affiliation{CAS Key Laboratory of Quantum Information, University of Science and Technology of China, Hefei 230026, People's Republic of China}
\affiliation{CAS Center for Excellence in Quantum Information and Quantum Physics, University of Science and Technology of China, Hefei 230026, People's Republic of China}

\author{Guang-Can Guo}
\affiliation{CAS Key Laboratory of Quantum Information, University of Science and Technology of China, Hefei 230026, People's Republic of China}
\affiliation{CAS Center for Excellence in Quantum Information and Quantum Physics, University of Science and Technology of China, Hefei 230026, People's Republic of China}

\begin{abstract}
The duration, strength and structure of memory effects are crucial properties of physical evolution. Due to the invasive nature of quantum measurement, such properties must be defined with respect to the probing instruments employed. Here, using a photonic platform, we experimentally demonstrate this necessity via two paradigmatic processes: future-history correlations in the first process can be erased by an intermediate quantum measurement; for the second process, a noisy classical measurement blocks the effect of history. We then apply memory truncation techniques to recover an efficient description that approximates expectation values for multi-time observables. Our proof-of-principle analysis paves the way for experiments concerning more general non-Markovian quantum processes and highlights where standard open systems techniques break down. 
\end{abstract}

\date{\today}
\maketitle

\textbf{Introduction.---}Memory effects are ubiquitous in nature~\cite{StochProc}, including disease spreading~\cite{Pimenov2012}, bio-chemical processes~\cite{Lambert2012,Scholes2017,Wang2019}, and optical fiber transmission~\cite{Banaszek2004}. Characterizing stochastic processes with memory is difficult because past events can impact the future, so long-term history must be recorded for accurate prediction. This necessitates developing memory truncation techniques for efficient approximation.

Stochastic processes arise from the inability to track all relevant degrees of freedom, which partitions the universe into an accessible \emph{system} and an inaccessible \emph{environment}, leading to open dynamics. A classical stochastic process on discrete times, $\{t_1, \dots, t_n \}$, is characterized by the $n$-point joint probability distribution over event sequences, $\mathbb{P}_n(x_1, t_1; \hdots; x_n,t_n)$. For a process with approximately finite-length memory, this distribution conditionally factors over any length-$\ell$ sequence of memory events $\{x_{k+1},\hdots,x_{k+\ell}\}$, with small error. This error is quantified by the conditional mutual information between the history $\{ x_1, \hdots, x_k\}$ and future $\{ x_{k+\ell+1}, \hdots, x_n\}$ events given the memory, which bounds the prediction accuracy when the history is truncated.

A key assumption here is that measurements do not affect the system. When invasive measurements are permitted in classical theory, such as in causal modeling~\cite{Pearl}, a joint probability distribution no longer describes the process~\cite{Milz2020KET}. Quantum theory is similar; however, here one cannot assume that non-invasive measurements could be made \emph{in principle}~\cite{piron_ideal_1981}, obfuscating the line between \emph{process} and \emph{observer}~\cite{Modi2011,Modi2012}. Many descriptions of open quantum dynamics have thus been restricted to two-time considerations~\cite{Rivas2014,Breuer2016}, where an operational picture of correlations between preparations and measurements arises via the dynamical map formalism~\cite{Sudarshan1961}. However, such approaches necessarily overlook multi-time correlations; these methods provide \emph{witnesses} of memory, but are insufficient to determine its presence~\cite{deVega2017,Li2018,Milz2017,Pollock2018L} or properties~\cite{Pollock2018A,Taranto2019L,Taranto2019A,Taranto2019S,TarantoThesis,Milz2019}. 

These issues have hindered the precise formulation of quantum stochastic processes and led to a `zoo' of definitions for memorylessness~\cite{Li2018}, some of which are contradictory~\cite{Mazzola2010,Haikka2011,Chruscinski2011}. Recently, the process tensor formalism~\cite{Pollock2018L,Pollock2018A,Milz2017}, which captures \emph{all} detectable multi-time correlations, has been developed (see also Refs.~\cite{Lindblad1979,Accardi1982,Kretschmann2005,Chiribella2008,Chiribella2009,Oreshkov2012,Oreshkov2016,Costa2016,Hardy2012,Hardy2016}). This framework separates the controllable impact on system dynamics due to an agent from the uncontrollable environmental influence: the former is described by quantum \emph{instruments}, which capture the post-measurement states for each (probabilistically-occurring) outcome; the process comprises the latter. This provides a consistent operational description of multi-time quantum stochastic processes that generalizes and unifies open quantum dynamics~\cite{Pollock2018A}. The process tensor correctly generalizes classical stochastic processes via a generalized Kolmogorov extension theorem~\cite{Milz2020KET}, and all memory properties such as Markovianity (memorylessness) and Markov order (finite-length memory) can be rigorously defined and recover classical definitions~\cite{Pollock2018L,Pollock2018A,Taranto2019L,Taranto2019A,Taranto2019S,Milz2020Classical}.

It was recently shown that there do not exist non-Markovian processes with finite-length memory for all instruments (although for any particular instrument, the memory length can be finite); thus operational descriptions of memory length must specify the probing instruments~\cite{Taranto2019L,Taranto2019A,Taranto2019S}. In this Letter, we experimentally demonstrate this instrument-specific nature of quantum memory via two three-time quantum processes on a photonic platform. Both processes are non-Markovian; however, by performing specific intermediate instruments, we show that future-history correlations can be deterministically erased, exemplifying finite-length memory for instruments. We then use memory truncation techniques developed in Ref.~\cite{Taranto2019S} to approximate non-Markovian processes with small memory strength and show this \emph{recovered} description to accurately predict multi-time expectation values. Our results provide the first demonstration of multi-time quantum memory effects beyond the two-time setting~\cite{bhliu2011,chiuri2012,zdliu2018} (see also Ref.~\cite{White2020} for a similar recent demonstration); while our proof-of-principle experiment focuses on ``common-cause processes''~\cite{Milz2020GMET,Nery2021}, in which correlations arise from an initial state, the methods employed are readily adaptable to the analysis of more general non-Markovian processes.

\textbf{Multi-time Quantum Processes.---}See the Supplemental Material \textbf{(SM)}~\cite{SM} for an introduction to the process tensor; here, we outline its key features. The process tensor is a linear mapping from sequences of quantum instruments---collections of completely-positive \textbf{(CP)} maps that sum to a completely-positive, trace-preserving \textbf{(TP)} map~\cite{Lindblad1979}---to the joint probability of their realization. Just as a density operator contains all necessary information to compute the probability of any measurement event via the Born rule, the process tensor encapsulates all information required to calculate the probability of realizing any sequence of events through a generalized Born rule~\cite{Shrapnel2018}. Any process tensor that decomposes into independent channels between timesteps is Markovian; by considering the distance to the nearest Markovian process, one can quantify the memory~\cite{Pollock2018L}. The process tensor can be tomographically reconstructed, constituting an operational description of quantum stochastic processes~\cite{Milz2018A}. Conversely, any operator satisfying generalized notions of complete-positivity and normalization, and a causality condition ensuring temporal order, represents some underlying open quantum dynamics~\cite{Pollock2018A,Chiribella2009}, i.e., can be dilated to a system-environment evolving unitarily between times, with the environment finally discarded. 

\begin{figure}[t!]
\centering
\includegraphics[width=0.95\linewidth]{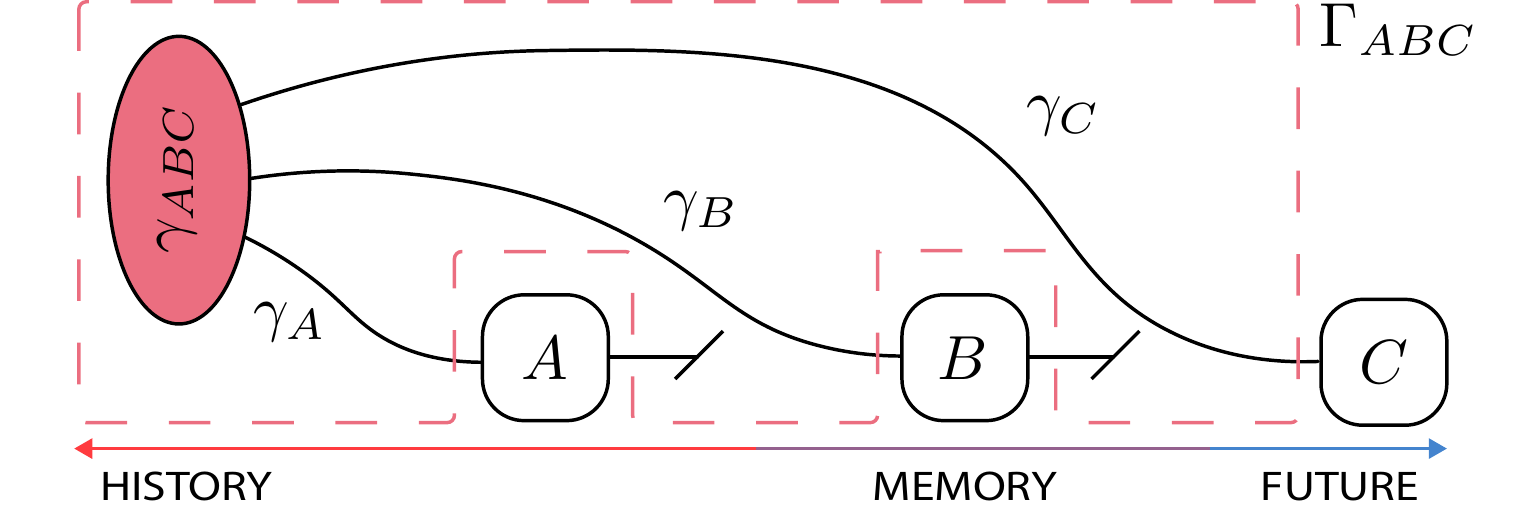}
\caption{\textit{Process Schematic.} Parts of an initial state $\gamma_{ABC}$ are sent to Alice, Bob and Charlie in a time-ordered fashion. Alice and Bob's outputs are discarded, corresponding to an identity matrix in the Choi representation, where each timestep has two Hilbert spaces (input and output). The resulting process tensor is $\Gamma_{ABC} = \gamma_{A^\inp B^\inp C^\inp} \otimes \mathbbm{1}_{A^\out B^\out}$. \label{fig:schematic}}
\end{figure}

While it is straightforward to compute the process tensor from a dilation, i.e., an underlying system-environment model (which is non-unique), it is difficult to engineer processes with certain memory properties. This is because correlations play a dynamical role and it is often unclear how to best design them within practical constraints. Moreover, the output states of each measurement generally influence future dynamics, presenting another experimental difficulty. To circumvent these problems, we examine two processes of a similar type, depicted in Fig.~\ref{fig:schematic}: ones for which subsystems of an initially-correlated state are fed out over time, with the output states discarded after each step. Such common-cause processes are a subset of general quantum processes that are amenable to current laboratory methods, as all correlations are encoded in the initial state. In particular, we examine the memory effects of two processes over three timesteps, which is the minimal setting for analyzing multi-time phenomena. We denote the initial three-qubit state of \emph{Process 1} by $\lambda_{ABC} \in \textsf{BL}(\mathcal{H}^2 \otimes \mathcal{H}^2 \otimes \mathcal{H}^2)$ and the corresponding process tensor by $\Lambda_{ABC}$. \emph{Process 2} is from Appendix E of Ref.~\cite{Taranto2019A}; we denote its initial qubit-qutrit-qubit state by $\omega_{ABC} \in \textsf{BL}(\mathcal{H}^2 \otimes \mathcal{H}^3 \otimes \mathcal{H}^2)$ and its process tensor by $\Omega_{ABC}$. Both common-cause states exhibit complicated correlations with non-trivial off-diagonal elements (see SM~\cite{SM}). The distinct memory effects displayed are due to the types of history-blocking instruments: for the former process, these are three-outcome qubit measurements with no classical analogue, demonstrating a genuinely quantum effect; for the latter process, these are noisy classical measurements, highlighting how coarse-graining can hide memory and positing Markovianity as an emergent phenomenon. 

\begin{figure*}
    \centering
    \includegraphics[width=0.7\linewidth]{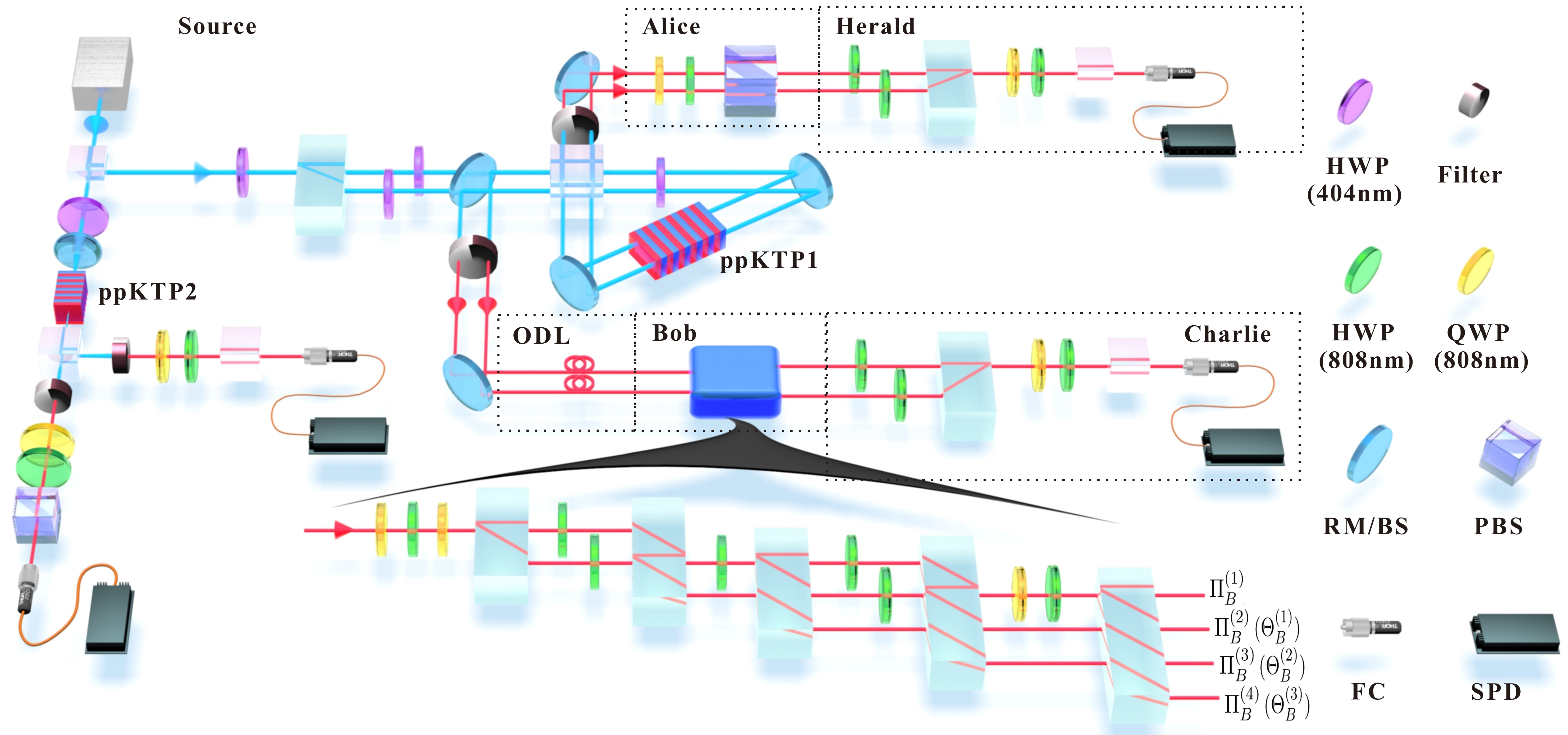}
    \caption{\emph{Experimental Setup.} An entangled photon pair is generated via spontaneous parametric down-conversion (SPDC) at ppKTP1 in a Sagnac interferometer pumped by a 404nm laser. For Process 1, the polarization acts as history and memory and path qubits as future and herald. For Process 2, an additional SPDC at ppKTP2 constructs a third level. Conditional Alice-Charlie correlations for each of Bob's outcomes encode memory effects. Inset shows Bob's measurement apparatus~\cite{Kurzynski2013,Zhao2015}. 
    }
    \label{fig:experimentalsetup}
\end{figure*}


\textbf{Experimental Setup.---}The processes considered comprise an initially-correlated state with parts sent out first to Alice (\emph{history}), then to Bob (\emph{memory}) and lastly to Charlie (\emph{future}), with each agent permitted freedom of choice and an optical delay line ensuring temporal order. The post-measurement states of Alice and Bob are discarded. The initial correlations encode the common-cause memory effects of the process~\cite{Costa2016,Ringbauer2017,ringbauer2018multi,Nery2021}; thus, state preparation is crucial to our experiment. We use a linear photonic system (see Fig.~\ref{fig:experimentalsetup}). The ``source'' prepares tripartite states encoded in path and polarization degrees of freedom of photon pairs. Although various techniques to construct multi-partite states exist~\cite{pan2012multiphoton,yokoyama2013optical,monz201114,ansmann2009violation}, many require distinct systems. We opt for a hybrid approach that encodes information in various degrees of freedom. This choice is motivated by the development of linear optical methods~\cite{takeuchi2000experimental,zhang2016engineering,reimer2019high}, which provide a high-fidelity and post-selection-free approach. The second critical element to our experiment is implementing positive operator-valued measurements \textbf{(POVMs)}. We use a discrete-time quantum walk protocol~\cite{Kurzynski2013,Zhao2015} (see SM~\cite{SM}).

\textbf{Results.---}\textit{Non-Markovianity.} For both processes $\Gamma_{ABC} \in \{ \Lambda_{ABC}, \Omega_{ABC}\}$, the non-Markovianity is the distance to its nearest Markovian counterpart
\begin{align}
    \mathcal{N} := \min_{\Gamma_{ABC}^{\textup{Markov}}} \mathcal{D}\left(\Gamma_{ABC}\|\Gamma_{ABC}^{\textup{Markov}}\right).
\end{align}
Choosing the quantum relative entropy $S(X\|Y):= \tr[X(\log X - \log Y)]$ as the distance~\footnote{Since entropies are only well-defined for normalized quantities, we calculate them on normalized process tensors in the Choi state representation, i.e., for $\hat{\Gamma} = \Gamma/\tr{[ \Gamma]}$, $S_X = -\hat{\Gamma}_X \log[\hat{\Gamma}_X]$.}, the minimum is achieved by the process constructed from its marginals~\cite{Vedral2002}, $\gamma_X := \ptr{YZ}[\gamma_{XYZ}]$, i.e., $\Gamma_{ABC}^{\textup{Markov}} = \gamma_{A^\inp} \otimes \mathbbm{1}_{A^\out} \otimes \gamma_{B^\inp} \otimes \mathbbm{1}_{B^\out} \otimes \gamma_{C^\inp}$. The following results hold for any CP-contractive (pseudo-)distance; the choice of the relative entropy bypasses optimization over Markovian processes and has an operational interpretation: $P_{\textup{confusion}} := \exp{(-n \mathcal{N})}$ is the probability of confusing the process with a promised Markovian one after $n$ measurements~\cite{Pollock2018L,Milz2020}. In the SM~\cite{SM}, we present the experimental tomographic data based on temporally-ordered measurements performed on the common-cause state at the three laboratories for both processes. For $\Lambda_{ABC}$, the non-Markovianity is $0.285\pm0.004$ and for $\Omega_{ABC}$ it is $0.458\pm0.004$, with theoretical predictions $0.329$ and $0.5$ respectively. 

Both processes are, however, CP-divisible, meaning all two-point dynamics can be described by composition of (fictitious) CPTP channels, i.e., $\Lambda_{A^\out C^\inp} = \Lambda_{B^\out C^\inp} \circ \Lambda_{A^\out B^\inp}$. As the output states discarded, one has $\Lambda_{A^\out B^\inp} = \mathbbm{1}_{A^\out} \otimes \gamma_{B^\inp}$ and $\Lambda_{B^\out C^\inp} = \mathbbm{1}_{B^\out} \otimes \gamma_{C^\inp}$; similarly, any common-cause process is CP-divisible, which is often used as a proxy for quantum Markovianity~\cite{Rivas2014,Breuer2016}. Nonetheless, CP-divisibility only considers two-point correlations; thus while it can witness non-Markovianity, it is insufficient to conclude that a process is Markovian, which requires checking multi-time conditions~\cite{Milz2019}.
For more general non-Markovian quantum processes than the common-cause ones considered here, all such two-point techniques necessarily fail, whereas the process tensor formalism is tailor-made for their analysis.

\textit{Markov Order.} We now demonstrate that both processes---although non-Markovian---have finite Markov order for particular instruments. The Markov order is the minimum number of times over which an agent must act to block history-future correlations~\footnote{Our methods apply to processes in which Bob could act multiple times between Alice and Charlie, capturing higher Markov order.}. Both processes have Markov order 1~\footnote{For quantum processes, Markov order 1 does not necessarily coincide with Markovianity: quantum Markovianity means that if Bob performs an informationally-complete measurement and independent repreparation, Charlie's state only depends upon Bob's repreparation, which is not the case here.}, meaning that Bob can apply an instrument $\mathcal{J}_B = \{ \mathsf{O}_B^{(x)}\}$ such that for each event, Alice and Charlie are conditionally independent
\begin{align}\label{eq:condprocess}
    \tr_B\left[\mathsf{O}_B^{(x) \textup{T}} \Gamma_{ABC}\right] = \Gamma_A^{(x)} \otimes \Gamma_C^{(x)}. 
\end{align}
By performing $\mathcal{J}_B$, Bob deterministically erases the memory. Deviation from Eq.~\eqref{eq:condprocess}, i.e., a correlated conditional process for any event of an instrument evidences longer memory with respect to said instrument. 

\begin{figure}[t]
    \centering
    \includegraphics[width=\linewidth]{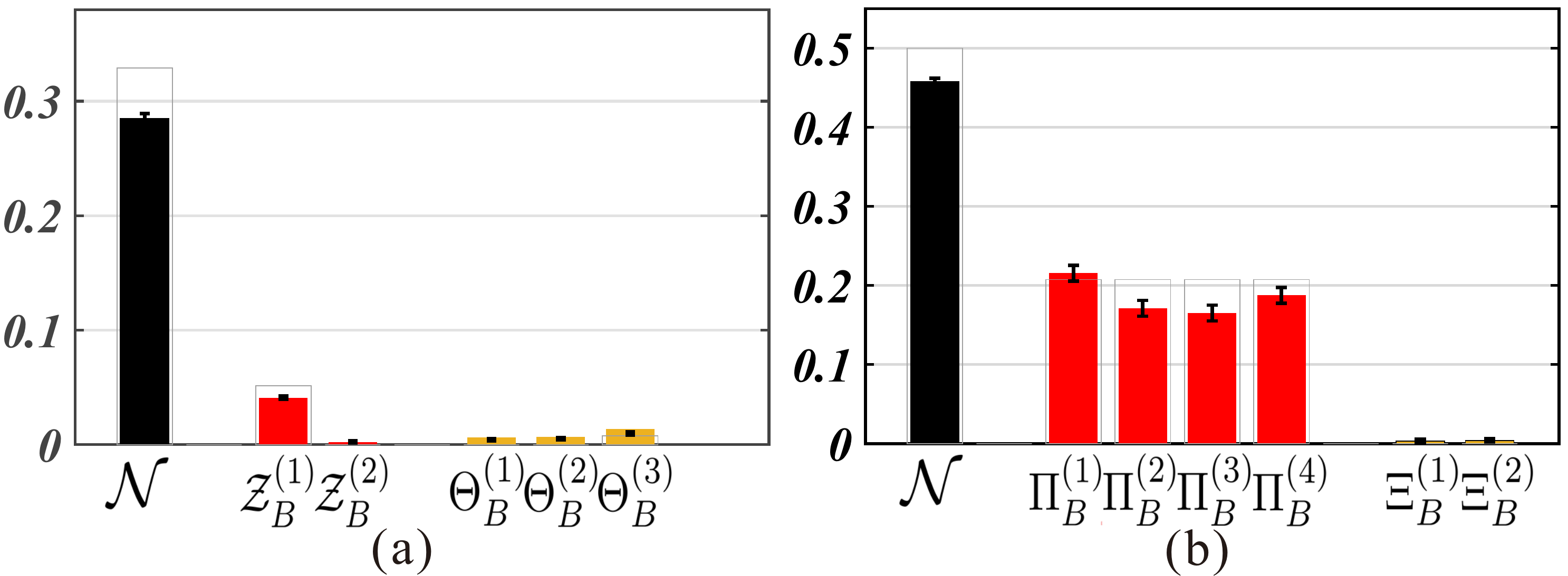}
    \caption{\emph{Non-Markovianity and memory strength of (a) Process 1 and (b) Process 2.} Experimental results for the non-Markovianity $\mathcal{N}$ (black, left), the (non-vanishing) memory strength for \emph{(a)} the computational-basis measurement $\mathcal{Z}_B = \{ \mathcal{Z}_B^{(x)} \}$ and \emph{(b)} POVM $\Pi_B = \{ \Pi_B^{(x)} \}$ (red, middle) (Eq.~\eqref{eq:tetrapovm}) and the (vanishing) memory strength for \emph{(a)} the POVM $\Theta_B = \{ \Theta_B^{(x)}\}$ (Eq.~\eqref{eq:perespovm}) and \emph{(b)} noisy measurement $\Xi_B = \{ \Xi_B^{(x)}\}$ (yellow, right). Each bar shows the mutual information between Alice and Charlie, conditioned on Bob's event. Gray edges show theoretical predictions. }
    \label{fig:results}
\end{figure}

For Process 1, $\Lambda_{ABC}$, the history-blocking instrument is a POVM $\Theta_B = \{ \Theta_B^{(x)}\}$ comprising  
\begin{align}\label{eq:perespovm}
    \Theta^{(1)}_B &= \frac{\sqrt{2}}{1+\sqrt{2}} \ket{1}\bra{1}, \notag \\ 
    \Theta_B^{(2)} &= \frac{\sqrt{2}}{2(1+\sqrt{2})} \left( \ket{0}-\ket{1} \right) \left( \bra{0}-\bra{1}\right), \notag \\ 
    \Theta_B^{(3)} &= \mathbbm{1}-\Theta^{(1)}_B-\Theta_B^{(2)}. 
\end{align}
Fig.~\ref{fig:results} (a) depicts the mutual information, $S_{AC} := S_A + S_C - S_{AC}$, with $S_X$ the von Neumann entropy, of the conditional processes. The memory strength for each event of $\Theta_B$ is $0.0042\pm 0.0010, 0.0053\pm 0.0010, 0.0098\pm 0.0014$, signifying negligible Alice-Charlie correlations, with an average memory strength of $(6.3\pm1.1)\times10^{-3}$. Conversely, if Bob measures in the computational basis $\mathcal{Z}_B = \{ \ket{0}\bra{0}, \ket{1}\bra{1}\}$ then Alice and Charlies' conditional processes are correlated; for instance, the first event of $\mathcal{Z}_B$ has memory strength $0.0410\pm0.0015$ [see middle bars in Fig.~\ref{fig:results} (a)]. The fact that this history-blocking instrument comprises a three-outcome POVM, which has no classical analogue, signifies that the approximate recovery we construct below represents a genuinely quantum approximation, in the sense that no recovery with respect to projective measurements would be as accurate or versatile. Nonetheless, for this process, certain projective measurements can render Alice and Charlie approximately conditionally-independent (see SM~\cite{SM}). We conjecture that in higher dimensions, there exist processes for which \emph{no} set of orthogonal projectors block the history, but certain POVMs do.

Process 2, $\Omega_{ABC}$, highlights how coarse-graining can block memory. If Bob performs a \emph{noisy} classical measurement $\Xi_B = \{ \Xi_B^{(x)} \} = \{ \mathbbm{1}_{01}, \ket{2}\bra{2}\}$ that cannot distinguish events on the first two levels of his qutrit, the process has Markov order 1, depicted by the rightmost bars in Fig.~\ref{fig:results} (b), which vanish for each event (the experimental value is $0.004\pm0.002$). On the other hand, if Bob performs a three-level measurement that resolves events in the first two levels, then Alice and Charlies' conditional processes can be correlated. For instance, consider the POVM $\Pi_B = \{ \Pi_B^{(x)}\}$:
\begin{align}\label{eq:tetrapovm}
    \Pi^{(x)}_B = \tfrac{1}{4} (\mathbbm{1}+\tfrac{1}{\sqrt{3}} \sum_j c_j^{(x)} \sigma_j),
\end{align}
where $\{ \mathbbm{1}, \sigma_X, \sigma_Y, \sigma_Z\}$ are Pauli matrices with coefficients $\{ c^{(x)} \} = \{ (1,1,1),(1,-1,-1),(-1,1,-1),(-1,-1,1)\}$. The measurement events respectively have memory strength $0.216\pm0.001, 0.171\pm0.0009, 0.165\pm0.001, 0.188\pm 0.0009$, as shown by the middle bars in Fig.~\ref{fig:results} (b) (see SM~\cite{SM}). These memory effects are close to maximal and one expects that some memory will be present for generic three-level POVMs; however, it is also likely that there exist fine-grained measurements that approximately render Alice and Charlie uncorrelated for each outcome. Our results are relevant towards understanding Markovianity as a consequence of coarse-graining. 

Knowledge of any approximately history-blocking instrument allows one to `recover' an efficient and accurate description, as we now demonstrate.

\begin{figure}[b!]
    \centering
    \includegraphics[width=1\linewidth]{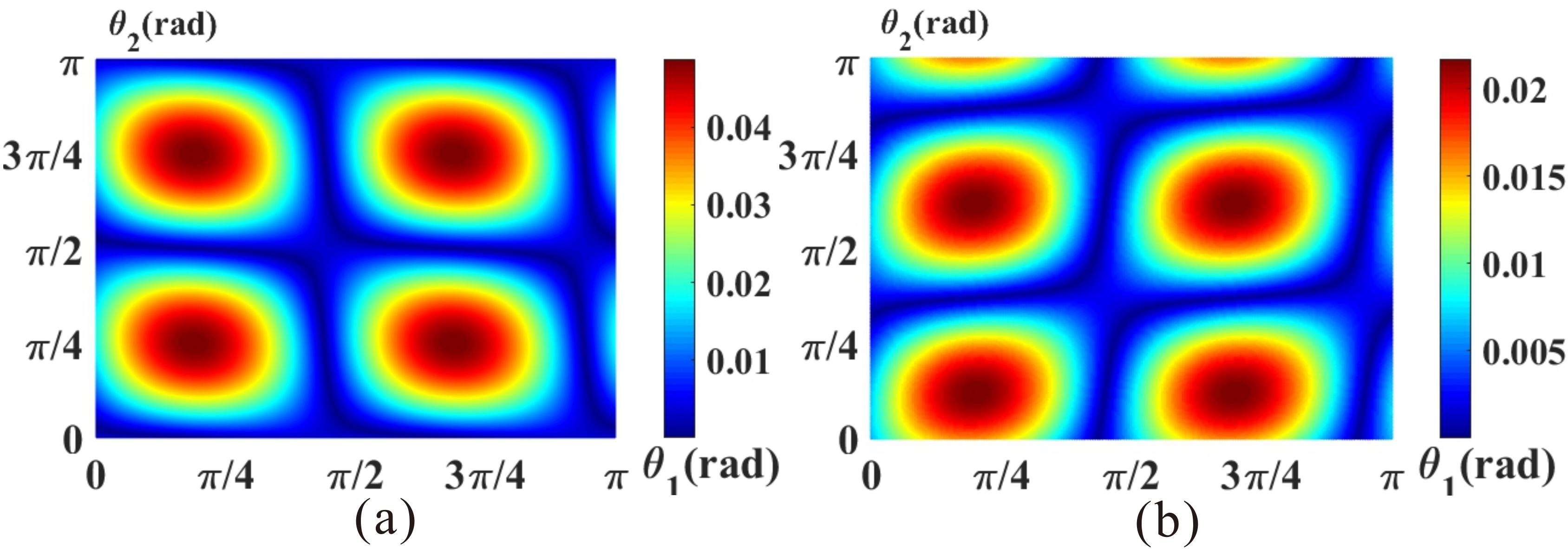}
    \caption{\emph{Multi-time expectations for (a) Process 1 and (b) Process 2.} Alice and Charlie project onto orthogonal states $\{|\phi\rangle=\cos{\theta_1}|0\rangle+e^{i \phi}\sin{\theta_1}|1\rangle, |\phi\rangle^\perp=\sin{\theta_1}|0\rangle-e^{-i \phi}\cos{\theta_1}|1\rangle\}$ and $\{|\psi\rangle=\cos{\theta_2}|0\rangle+e^{i \psi}\sin{\theta_2}|1\rangle, |\psi\rangle^\perp=\sin{\theta_2}|0\rangle - e^{-i \psi}\cos{\theta_2}|1\rangle\}$ respectively and Bob performs any non-selective measurement. The difference of expectation values via the recovered and true process is plotted for fixed phase \emph{(a)} $\phi=0$ and $\psi=0$ and \emph{(b)} $\phi=1.920\pi$ and $\psi=\pi$. }
    \label{fig:simulation}
\end{figure}

\textit{Efficient Recovery.} In general, the conditional processes for an instrument applied by Bob are correlated. Aggregating the mutual information of the conditional processes to the instrument level quantifies the memory strength in an instrument-specific manner~\cite{Taranto2019S}. One can then upper bound the difference between a class of multi-time expectation values calculated with the actual process (i.e., with common-cause state $\gamma_{ABC}$) versus a \emph{recovered} process (see below), which efficiently approximates the true one by discarding future-history correlations (see SM~\cite{SM}).

Since both processes have vanishing memory strength for some instrument, one can reconstruct an accurate such recovered processes. For common-cause processes, the reconstruction with respect to Bob's history-blocking measurement $\mathcal{J}_{B^\inp} = \{ \mathsf{O}_{B^\inp}^{(x)}\}$ has the common-cause state:
\begin{align}\label{eq:mainreconstruct}
    \underline{\gamma}_{ABC}^{\mathcal{J}_B} = \sum_x \gamma_{A^\inp}^{(x)} \otimes \Delta_{B^\inp}^{(x)} \otimes \gamma_{C^\inp}^{(x)},
\end{align}
where $\{ \gamma_{A^\inp}^{(x)} , \gamma_{C^\inp}^{(x)} \}$ are the marginals of Alice and Charlie for each outcome Bob yields and $\{ \Delta_{B^\inp}^{(x)} \}$ satisfies $\tr{\left[\Delta_{B^\inp}^{(x)} \mathsf{O}_{B^\inp}^{(y) \textup{T}}\right]} = \delta_{xy}$ (such a set of operators always exists for linearly-independent measurement elements~\cite{Modi2012A}). We tomographically reconstruct the marginals from local measurements of Alice and Charlie for Bob's outcomes. The approximate process $\underline{\Gamma}_{ABC}^{\mathcal{J}_B} = \underline{\gamma}_{ABC}^{\mathcal{J}_B} \otimes \mathbbm{1}_{A^\out B^\out}$ accurately predicts expectation values for any observable of the form $C_{ABC} = \sum_{y} c_y \mathsf{X}^{(y)}_{ABC}$ where $\mathsf{X}^{(y)} = \sum_{x} \! \mathsf{E}^{(x,y)}_{AC} \!\otimes \mathsf{O}^{(x)}_{B}$, with arbitrary $\mathsf{E}^{(x,y)}_{AC}$; this ensures that Bob's observable is in the span of the original instrument, as required for the recovered process to make sensible predictions~\cite{Taranto2019S}. 

In Fig.~\ref{fig:simulation} we consider projective measurements for Alice and Charlie, with Bob performing any non-selective measurement. This choice permits arbitrary product Alice-Charlie observables; more complicated temporally-correlated observables would require higher levels of control that are not presently available. We scan the parameter space and compare the expectation values calculated from the recovered and true process, finding maximum deviations of $0.048$ and $0.022$ for Processes 1 and 2 respectively; the maximum difference for an arbitrary process is 1, highlighting the approximation accuracy. Lastly, the memory strength with respect to Bob's instrument $\mathcal{J}_B$ bounds the inaccuracy of approximating \emph{any} observable of the form below Eq.~\eqref{eq:mainreconstruct}; this includes, e.g., correlated Alice-Charlie observables~\cite{Taranto2019S}. Calculating the memory strength for $\Theta_B$ and $\Xi_B$ for the two processes respectively gives a bound of $0.450\pm0.017$ and $0.348\pm0.015$. 

\FloatBarrier

\textbf{Conclusions \& Outlook.---}\label{sec:discussion}In this Letter, we have demonstrated the instrument-specific nature of quantum memory. Our experiment provides the first report of finite quantum Markov order for non-Markovian common-cause processes, i.e., the ability to erase future-history correlations, stored in a correlated initial state, via a specific instrument. This has implications for the approximation of quantum processes with memory, which we highlighted by reconstructing the recovered process that disregards negligible correlations and showing this to accurately predict multi-time expectation values. Such memory truncation techniques are pivotal to efficiently characterizing near-term quantum devices.

Our work opens some important avenues: we have posed the question of whether there exist processes for which no orthogonal measurement blocks the history, but certain POVMs do; and that concerning the emergence of Markovianity via coarse-graining. Analyzing the relative volumes of measurement space that lead to finite Markov order, as well as the tightness of the bounds, is warranted to develop more robust approximations. 

The common-cause processes analyzed here are an important class of non-Markovian processes; in particular, they display global memory properties that could not be characterised from standard two-point measurement techniques. On the other hand, all correlations arising from such processes can be computed from measurements on the initial common-cause state, as the post-measurement states play no role, making them more amenable to current experimental platforms. By phrasing the analysis of such memory effects in the operational process tensor formalism and highlighting where other techniques would fail, our work provides a starting point for the analysis of more general processes (e.g., those for which post-measurement states play a role). A pertinent example to this end is the recent experiment performed in Ref.~\cite{White2020}, which characterised the memory effects of a non-Markovian process on the span of unitary operations. Such experiments will require high levels of quantum control and the ability to implement multi-input to multi-output gates. As such technical challenges are overcome, a holistic analysis of multi-time quantum memory effects will become possible.

\FloatBarrier

\begin{acknowledgments}
We thank Kavan Modi for discussions. This work was supported by the National Key Research and Development Program of China (No.\ 2017YFA0304100, No.\ 2016YFA0301300 and No.\ 2016YFA0301700), NSFC (Nos.\ 11774335,\ 11734015,\ 11874345,\ 11821404,\ 11904357), the Key Research Program of Frontier Sciences, CAS (No.\ QYZDY-SSW-SLH003), Science Foundation of the CAS (No.\ ZDRW-XH-2019-1), the Fundamental Research Funds for the Central Universities, and Anhui Initiative in Quantum Information Technologies (Nos.\ AHY020100, AHY060300). P.T. was supported by the Austrian Science Fund (FWF) START project (No.\ Y879-N27).
\end{acknowledgments}

\FloatBarrier

%

\FloatBarrier

\appendix 
\section*{Supplemental Material}

\section{Process Tensor Formalism}\label{app:processtensor}

\subsection{Introduction to Process Tensor}

Here we provide a brief introduction to the process tensor formalism used throughout the main text; for a more thorough introduction, see, e.g., Refs.~\cite{Chiribella2009,Pollock2018A,Milz2017}. 

A discrete-time classical stochastic process is completely described by the joint probability distribution $\mathbbm{P}$ that associates the correct probabilities to all sequences of events $\{ x_1, \hdots, x_n \}$ at the times specified $\{ t_1, \hdots, t_n \}$; this is altogether denoted by $\mathbbm{P}( x_n, \hdots, x_1)$, where we drop the explicit time labels with the understanding that $x_j$ represents an event occurring at time $t_j$. In quantum mechanics, it is important to not only capture the outcome of a measurement, but also the transformation induced on the state via the measurement, as the post-measurement goes on to influence the future dynamics; together, these correspond to the notion of an \emph{event} in quantum mechanics. Thus, an interrogation of a quantum stochastic process at some time $t_j$ is described by an instrument $\mathcal{J}_j = \{ \mathsf{O}_j^{(x_j)} \}$, which is a collection of completely positive (CP) maps that sum to a completely positive and trace preserving (CPTP) map. Intuitively, each map in the collection corresponds to a particular event realized by the experimenter upon probing the process; the fact that the maps sum to a CPTP one encode the assumption that the experimenter measures \emph{some} event, thereby implementing an instrument with overall certainty. A discrete-time quantum stochastic process is uniquely described once all of the probabilities $\mathbbm{P}(x_n, \hdots, x_1 | \mathcal{J}_n, \hdots, \mathcal{J}_1)$ for all possible sequences of events $\{ x_n, \hdots, x_1 \}$ for all possible instrument sequences $\{ \mathcal{J}_n, \hdots, \mathcal{J}_1\}$ at probing times $\{ t_n, \hdots, t_1\}$ are known. Due to the probabilistic nature of quantum mechanics---where the linearity of mixing principle must hold---there exists a multi-linear functional that takes any sequence of CP maps to the correct probability of their realization; this object is called the \emph{process tensor}. 

\begin{figure}[t!]
    \centering
    \includegraphics[width=\linewidth]{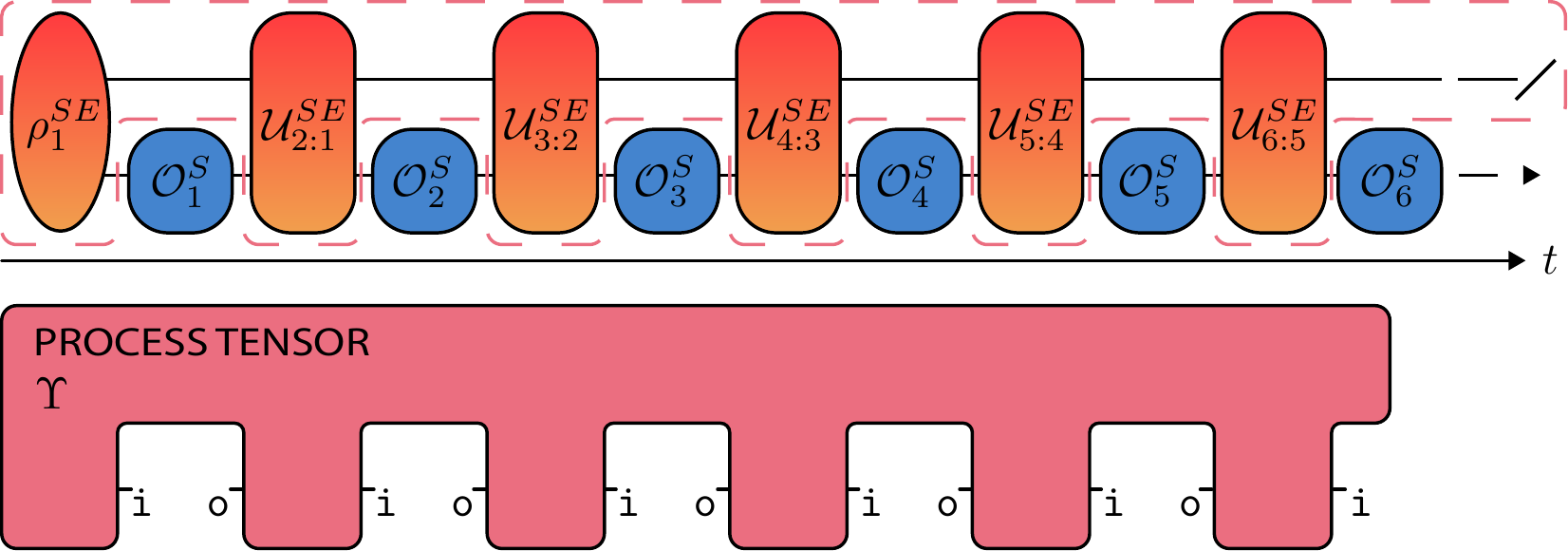}
    \caption{\textit{Open system dynamics and process tensor.} An open quantum process with the system being probed at finite times throughout its evolution. By abstracting everything that is out of control of the experimenter (within the dashed line and shown below), one yields a process tensor. The process tensor is isomorphic to a quantum state satisfying a hierarchy of causality conditions (see Eq.~\eqref{eq:causalityhierarchy}); conversely, any operator satisfying these conditions corresponds to some open dynamics.} 
    \label{fig:dilation}
\end{figure}

Since all of the CP maps constituting the instruments, as well as the process tensor itself, are linear maps, we can represent all of these objects as matrices through the Choi-Jamio{\l}kowski isomorphism~\cite{Milz2017}: any map $\mathcal{O}: \mathcal{B}(\mathcal{H}_\inp) \to \mathcal{B}(\mathcal{H}_\out)$ can be mapped isomorphically to a matrix $\mathsf{O} \in \mathcal{B}(\mathcal{H}_\out \otimes \mathcal{H}_\inp)$ by letting it act on half of an (unnormalized) maximally entangled state $\Phi^+ = \sum_{i,j} \ket{ii}\bra{jj} $, i.e., $\mathsf{O} := (\mathcal{O} \otimes \mathcal{I})[\Phi^+]$. Similarly, the process tensor itself can be represented as $\Upsilon_{n:1} \in \mathcal{B}(\mathcal{H}_{n^\inp} \otimes \mathcal{H}_{n-1^\out} \otimes \hdots \otimes \mathcal{H}_{1^\inp})$. Note that to each timestep of a quantum stochastic process is associated an input and an output Hilbert space, which we label logically from the perspective of the experimenter (i.e., the experimenter receives a state from the process that is ``input'' into their instrument of choice, transforming it into an ``output'' state that is fed back into the process). From these objects, the correct probabilities are calculated via the generalised temporal Born rule~\cite{Shrapnel2018}:
\begin{align}\label{eq:appgbornrule}
    &\mathbbm{P}(x_n, \hdots, x_1 | \mathcal{J}_n, \hdots, \mathcal{J}_1) \notag \\
    &= \tr{\left[(\mathsf{O}_n^{(x_n)} \otimes \hdots \otimes \mathsf{O}_1^{(x_1)} )^\textup{T} \Upsilon_{n:1} \right]}. 
\end{align}
In the Choi representation, natural generalizations of the notions of complete positivity and trace preservation translate respectively to the following properties of the process tensor: $\Upsilon_{n:1} \geq 0$ and $\tr{[\Upsilon_{n:1}]} = d^n$, where $d$ is the dimension of the system of interest. Moreover, imposing causality, i.e., that no signal can be sent from the future to the past on average, implies the following hierarchy of trace conditions:
\begin{align}\label{eq:causalityhierarchy}
    \ptr{j^\inp}{[\Upsilon_{j:1}]} = \mathbbm{1}_{j-1^\out} \otimes \Upsilon_{j-1:1}, \quad \forall j.
\end{align}
Conversely, any matrix satisfying the above properties represents some valid quantum stochastic process~\cite{Chiribella2009,Pollock2018A}; the (non-unique) dilation of a quantum stochastic process is shown in Fig.~\ref{fig:dilation}. 

By restricting the trace of Eq.~\eqref{eq:appgbornrule} to a subset of timesteps over which an instrument sequence has been specified, the conditional process defined on the remaining timesteps can be calculated. For instance, grouping the times into history $\{ t_1, \hdots, t_{k}\}$, memory $\{ t_{k+1}, \hdots , t_{k+\ell}\}$ and future $\{ k_{k+\ell+1}, \hdots, t_n\}$ and choosing an instrument $\mathcal{J}_M = \{ \mathsf{O}_M^{(x_M)} \}$ on the memory block, then the conditional future-history process that occurs given any particular event $\mathsf{O}_M^{(x_M)}$ is
\begin{align}
    \Upsilon_{FH}^{(x_M)} = \ptr{M}{\left[\mathsf{O}_M^{(x_M) \textup{T}} \Upsilon_{FMH}\right]}.
\end{align}
Such a conditional process is generically correlated; however, if it is of tensor product form $\Upsilon_{FH}^{(x_M)} = \Upsilon_F^{(x_M)} \otimes \Upsilon_H^{(x_M)}$ for each event $x_M$ of the instrument $\mathcal{J}_M$, the process has Markov order $\ell := |M|$ with respect to said instrument~\cite{Taranto2019L}.

Lastly, a Markovian process corresponds to one for which the process tensor has the specific tensor product structure of an uncorrelated sequence of CPTP maps connecting adjacent timesteps, plus an initial quantum state of the system~\cite{Pollock2018L}:
\begin{align}
    \Upsilon_{n:1} = \Lambda_{n^\inp:n-1^\out} \otimes \Lambda_{n-1^\inp:n-2^\out} \otimes \Lambda_{2^\inp:1^\out} \otimes \rho_{1^\inp}. 
\end{align}

\subsection{Theoretical Details of Processes Constructed}

Both of the processes that we consider are defined over three timesteps, labeled by Alice (History), Bob (Memory) and Charlie (Future), and follow the general outline depicted in Fig.~1 of the main text and described as follows. 

In each case, the process prepares a particular tripartite quantum state, $\gamma_{ABC}$, each part of which is fed out to the parties sequentially in time. Alice, upon receiving her part, can apply any instrument that she likes; however, the process simply discards her output state and feeds to Bob the second part of the initial state. Similarly, Bob can implement any instrument of his choosing, but the process discards his output state and sends Charlie the third part of the initial state, upon which he can perform any measurement. Since the output states of Alice and Bob are discarded, we can obtain the process tensor through the tensor product of the initial tripartite state on all of the input spaces $\gamma_{A^\inp B^\inp C^\inp}$ with identity matrices on the output spaces, i.e., $\Gamma_{ABC} = \gamma_{A^\inp B^\inp C^\inp} \otimes \mathbbm{1}_{A^\out B^\out}$. The only difference between the two processes considered is that different initial states are prepared by the process. In both cases, it is clear to see that the processes are CP-divisible but non-Markovian~\cite{Milz2019}. Furthermore, both processes have non-vanishing quantum conditional mutual information (CMI) between $A$ and $C$ given $B$, i.e., $I(A:C|B) := S_{AB} + S_{BC} - S_{ABC} - S_B \neq 0$; this provides yet another distinction to Markov order in the standard classical case, where the CMI necessarily vanishes.

\emph{Process 1.} The following tripartite state is constructed: 
\begin{widetext}
\begin{align}
    \lambda_{ABC} = \frac{1}{10000} \left[ \begin{array}{cccccccc} 1106 & 25 & -142 & -525 & 25 & 294 & -525 & -58 \\
    25 & 1106 & -525 & -142 & 294 & 25 & -58 & -525 \\
    -142 & -525 & 1394 & 25 & -525 & -58 & 25 & 6 \\
    -525 & -142 & 25 & 1394 & -58 & -525 & 6 & 25 \\
    25 & 294 & -525 & -58 & 1106 & 25 & -142 & -525 \\
    294 & 25 & -58 & -525 & 25 & 1106 & -525 & -142 \\
    -525 & -58 & 25 & 6 & -142 & -525 & 1394 & 25 \\
    -58 & -525 & 6 & 25 & -525 & -142 & 25 & 1394 \end{array} \right].
\end{align}
\end{widetext}

The process proceeds as described above, thus the process tensor is $\Lambda_{ABC} = \lambda_{A^\inp B^\inp C^\inp} \otimes \mathbbm{1}_{A^\out B^\out}$. As discussed in the main text, the process is non-Markovian; however, Bob can implement a specific POVM such that for each event that he observes, Alice and Charlie are rendered (approximately) independent. The particular POVM $\Theta_B$ comprises elements
\begin{align}\label{eq:appperespovm}
    \Theta^{(1)}_B &= \frac{\sqrt{2}}{1+\sqrt{2}} \ket{1}\bra{1}, \notag \\ 
    \Theta_B^{(2)} &= \frac{\sqrt{2}}{2(1+\sqrt{2})} \left( \ket{0}-\ket{1} \right) \left( \bra{0}-\bra{1}\right), \notag \\ 
    \Theta_B^{(3)} &= \mathbbm{1}-\Theta^{(1)}_B-\Theta_B^{(2)}. 
\end{align}
Conditioned on each event $x$, which occur with probabilities $\{2(3-2\sqrt{2}),2(3-2\sqrt{2}),8\sqrt{2}-11\}$, the joint Alice-Charlie process factorizes: 
\begin{align}
    \Lambda_{AC}^{(x)} &= \ptr{B}{\left[ \left(\Theta_{B^\inp}^{(x)} \otimes \mathbbm{1}_{B^\out}\right)^\textup{T} \Lambda_{ABC} \right]} \notag \\
    &\approx \lambda_{A^\inp}^{(x)} \otimes \mathbbm{1}_{A^\out} \otimes \lambda_{C^\inp}^{(x)},
\end{align}
with $\{ \lambda_X^{(x)} \}$:
\begin{align}\label{eq:applambda}
\lambda_X^{(1)}&={
\left[ \begin{array}{cc}
0.5000&0.008967\\
0.008967&0.5000\\
\end{array}
\right ]}, \notag \\
\lambda_X^{(2)}&={
\left[ \begin{array}{cc}
0.5000&0.1976\\
0.1976&0.5000\\
\end{array}
\right ]}, \notag\\
\lambda_X^{(3)}&={
\left[ \begin{array}{cc}
0.5000&-0.1652\\
-0.1652&0.5000\\
\end{array}
\right ]}.
\end{align}
with $X \in \{ A^\inp , C^\inp \}$. In particular, there are approximately no $AC$ correlations in any of the conditional processes, and since the process must be conditioned into one of them (depending on the event), Bob's instrument serves to deterministically block any correlation from the history to the future. As such, the process is said to approximately have Markov order 1. However, if he chooses any other instrument, significant $AC$ correlations generally exist in the conditional processes (at least for some events). For example, if Bob measures in the computational basis $\mathcal{Z}_B = \{ \ket{0}\bra{0}, \ket{1}\bra{1}\}$ then Alice and Charlies' conditional process for the outcome $x=0$ is
\begin{align}
\lambda_{A^\inp C^\inp}^{(0)}&={
\left[ \begin{array}{cccc}
0.2500&0.005651&0.005651&0.06646\\
0.005651&0.2500&0.06646&0.005651\\
0.005651&0.06646&0.2500&0.005651\\
0.06646&0.005651&0.005651&0.2500\\
\end{array}
\right ]},
\end{align}
which is correlated with memory strength 0.0514. Thus, the process displays memory effects for generic instruments. 

Importantly, the POVM elements of Bob's measurement are non-orthogonal, so this particular  history-blocking instrument has no classical counterpart. This implies that the recovery process that we approximate (see below) with respect to said instrument is genuinely quantum and could not be reconstructed using knowledge of any classical interrogations. Nonetheless, for this process there do exist orthogonal projective measurements that block the history: choosing a cutoff for the maximum memory strength over outcomes---i.e., the worst-case scenario---as 0.0125, which accounts for two standard errors from the maximal memory strength of the POVM $\Theta_B$ in our experimental data (see main text), we find that $\approx 28.8\%$ of projective measurements yield a memory strength below the cutoff value; hence the majority of projective measurements can be reliably distinguished from the non-orthogonal history-blocking instrument  $\Theta_B$ with experimental confidence. While one could reconstruct a recovered process for any such projective instrument that effectively blocks the history, it would be a less accurate approximation of the true process than that recovered with respect to the more fine-grained POVM, which necessarily accounts for more information; moreover, the set of observables it could be used to accurately approximate would be more limited. In addition, as the relative proportion of measurements that block the history are expected to shrink for higher-dimensional quantum systems, we conjecture that there exist processes for which \emph{no} set of orthogonal projectors blocks the history, but certain POVMs do; such processes could be said to display \emph{genuinely} quantum memory effects.

Lastly, as a brief aside, note the quantum CMI of the process tensor does not vanish $I(A:C|B) \approx 0.019$, reflecting the result of Prop.~5 in Ref.~\cite{Taranto2019L}, which posits the existence of quantum processes with finite Markov order which nonetheless do not have vanishing quantum CMI.

\emph{Process 2.} The process begins with the following initial tripartite state:
\begin{widetext}
\begin{align}
    \omega_{ABC} = \frac{1}{48} \left[ \begin{array}{cccccccccccc} 3 & \sqrt{3} & 0 & 0 & 0 & 0 & 0 & 0 & \sqrt{3} & -\sqrt{3} i & 0 & 0 \\
    \sqrt{3} & 3 & 0 & 0 & 0 & 0 & 0 & 0 & \sqrt{3} i & -\sqrt{3} & 0 & 0 \\
    0 & 0 & 3 & -\sqrt{3} & 0 & 0 & -\sqrt{3} & -\sqrt{3} i & 0 & 0 & 0 & 0 \\
    0 & 0 & -\sqrt{3} & 3 & 0 & 0 & \sqrt{3} i & \sqrt{3} & 0 & 0 & 0 & 0 \\
    0 & 0 & 0 & 0 & 24 & 0 & 0 & 0 & 0 & 0 & 0 & 0 \\
    0 & 0 & 0 & 0 & 0 & 0 & 0 & 0 & 0 & 0 & 0 & 0 \\
    0 & 0 & -\sqrt{3} & -\sqrt{3} i & 0 & 0 & 3 & -\sqrt{3} & 0 & 0 & 0 & 0 \\
    0 & 0 & \sqrt{3} i & \sqrt{3} & 0 & 0 & -\sqrt{3} & 3 & 0 & 0 & 0 & 0 \\
    \sqrt{3} & -\sqrt{3} i & 0 & 0 & 0 & 0 & 0 & 0 & 3 & \sqrt{3} & 0 & 0 \\
    \sqrt{3} i & -\sqrt{3} & 0 & 0 & 0 & 0 & 0 & 0 & \sqrt{3} & 3 & 0 & 0 \\
    0 & 0 & 0 & 0 & 0 & 0 & 0 & 0 & 0 & 0 & 0 & 0 \\
    0 & 0 & 0 & 0 & 0 & 0 & 0 & 0 & 0 & 0 & 0 & 0 \end{array} \right].
\end{align}
\end{widetext}
The process tensor is $\Omega_{ABC} = \omega_{A^\inp B^\inp C^\inp} \otimes \mathbbm{1}_{A^\out B^\out}$.

Now, consider the instrument made up of the following two noisy, orthogonal operations: 
\begin{align}\label{eq:appfuzzy}
    &\Xi^{(1)}_{2^\inp} = (\mathbbm{1} - \ket{2}\bra{2})_{2^\inp} \notag \\
    &\Xi^{(2)}_{2^\inp} = \ket{2}\bra{2}_{2^\inp}.
\end{align} 
With respect to this instrument, the conditional process tensors for each event are:
\begin{align}\label{eq:condptcondwerner}
    \Omega^{(1)}_{AC } &= \frac{\mathbbm{1}_{A^\inp}}{2} \otimes \mathbbm{1}_{A^\out} \otimes \frac{\mathbbm{1}_{C^\inp}}{2} \notag \\
    \Omega^{(2)}_{AC} &= |0\rangle\langle 0|_{A^\inp} \otimes \mathbbm{1}_{A^\out} \otimes |0\rangle\langle 0|_{C^\inp}. 
\end{align}
Thus, the process has Markov order $1$ with respect to this instrument comprising only (noisy) orthogonal projectors. 

However, suppose now that Bob is able to resolve measurements in the $\{ \ket{0}, \ket{1} \}$ subspace of his qutrit, e.g., apply the instrument comprising the operations 
\begin{align}\label{eq:appqutritsharp}
    &\mathsf{O}_{2^\inp}^{(x)} = \Pi^{(x)}_{2^\inp} \quad \text{for } x \in \{ 1,2,3,4 \} \notag \\
    &\mathsf{O}^{(5)}_{2^\inp} = \ket{2}\bra{2}_{2^\inp},
\end{align}
with the POVM $\{ \Pi^{(x)}_{2^\inp} \}$ defined as follows:
\begin{align}\label{eq:apptetrapovm}
    \Pi^{(x)} = \tfrac{1}{4} (\mathbbm{1}+\tfrac{1}{\sqrt{3}} \sum_j c_j^{(x)} \sigma_j),
\end{align}
where $\{ \mathbbm{1}, \sigma_X, \sigma_Y, \sigma_Z\}$ are Pauli matrices with coefficients $\{ c^{(x)} \} = \{ (1,1,1),(1,-1,-1),(-1,1,-1),(-1,-1,1)\}$. Then the conditional process tensors for each event are:
\begin{align}\label{eq:condptcondwernercorr}
    &\Omega^{(x)}_{A^\inp A^\out C^\inp} = \psi^{(x)}_{A^\inp C^\inp} \otimes \mathbbm{1}_{A^\out} \notag \\
    &\Omega^{(5)}_{A^\inp A^\out C^\inp} = \ket{0}\bra{0}_{A^\inp} \otimes \mathbbm{1}_{A^\out} \otimes \ket{0}\bra{0}_{C^\inp}, 
\end{align}
with $\{\psi^{(x)} \}$ the qubit Werner states defined in terms of a probabilistic mixture between the maximally-mixed state and each of the four Bell pairs [see Eqs.~\eqref{eq:appwerner},~\eqref{eq:appbell}]. Although the event corresponding to level 2 of Bob's qutrit, which occurs with probability $q = \tfrac{1}{2}$, renders Alice and Charlie independent, this is not the case for each event of the overall instrument. For each event $x$ observed in the $\{ \ket{0}, \ket{1} \}$ subspace, the conditional Alice-Charlie processes exhibit correlations via one of the four Werner states with mixing parameter $r = \tfrac{1}{3}$ (which are separable, but not product, and therefore classically correlated): 
\begin{align}\label{eq:appwerner}
	\psi^{(x)}_{A^\inp C^\inp} = r \beta^{(x)} + (1-r) \frac{\mathbbm{1}}{2},
\end{align}
where each $\beta^{(x)}$ is the projector of one of the Bell pairs:
\begin{align}\label{eq:appbell}
	&\ket{\psi^\pm} := (\ket{00}\pm\ket{11})/\sqrt{2}, \notag \\
	&\ket{\phi^\pm} := (\ket{01}\pm\ket{10})/\sqrt{2}. 
\end{align}
Again, although the process has Markov order 1 for the noisy measurement, a straightforward calculation shows that the quantum CMI does not vanish: $I(A:C|B) = \tfrac{1}{2}$. 

\section{Experimental Details}\label{app:experimentaldetails}

\subsection{General Experimental Design}

We utilize a linear photonic system in our experimental investigation of quantum Markov order. The full setup is illustrated in Fig.~2 of the main text and can be divided into four modules: state preparation module (\emph{Source}), Alice's module, Bob's module, and Charlie's module. Arbitrary 4-qubit pure states can be prepared in the source module via spontaneous parametric down-conversion (SPDC) at a type II cut periodically poled potassium titanyl phosphate (ppKTP1) crystal~\cite{hu2018beating,guo2019experimental}. The first two qubits (acting as the history state [sent to Alice] and the memory state [sent to Bob] respectively) are encoded in the polarization ($|H\rangle$ for horizontal and $|V\rangle$ for vertical) of the two photons generated in the SPDC, while the third and fourth qubits are encoded in the latitudinal spatial mode of the photons (acting as a herald and the future state [sent to Charlie] respectively). Under the combined action of half-wave plates (HWPs), quarter-wave plates (QWPs) (not shown in Fig.~2) and a beam displacer (BD), the photon from a laser (@404nm) is prepared in the state
\begin{align}
    (\alpha|H\rangle+\beta|V\rangle)|0\rangle+(\gamma|H\rangle+\delta|V\rangle)|1\rangle.
\end{align}
After the SPDC process at ppKTP1, a three qubit state of the following form is generated
\begin{align}
    |\gamma \rangle=(\alpha|HV\rangle+\beta|VH\rangle)|0\rangle+(\gamma|HV\rangle+\delta|VH\rangle)|1\rangle,
\end{align}
where the complex coefficients can be adjusted by setting the angles of half-wave and quarter-wave plates. After projecting the herald qubit on the diagonal basis, the resulting states are exactly the form of the components of the needed ensemble [Eq.~\eqref{ensemble1}] of Process 1 and also the first four states of ensemble [Eq.~\eqref{ensemble2}] of Process 2. For the last state of ensemble in Eq.~\eqref{ensemble2}, a third-level basis of Bob's state is needed and is constructed with an additional SPDC (at ppKTP2) where the pump photon $|H\rangle|2\rangle$ splits into $|H\rangle|V\rangle|2\rangle$ through SPDC process. So, at each round of the experiment, we can randomly prepare one of the states from either ensemble with a corresponding probability, resulting in the required initial state $\gamma_{ABC}$ or $\omega_{ABC}$ over numerous rounds. 

The participants operate on their qubits (or qutrit for Bob in Process 2) in succession: the temporal order of the participants' operation is guaranteed by an optical time delay. Another key point is that both Alice's and Bob's post-measurement states cannot be fed forward in the process. This is ensured by the fact that the qubits (or qutrits) are encoded in different subspaces or on different photons. As such, Charlie’s measurement apparatus has no access to Alice's or Bob's qubit (or qutrit) subspace and so cannot perform any information processing on them; similarly, Bob cannot access Alice's post-measurement qubit.

\subsection{State Preparation}
The eigendecomposition of the initial mixed state in Process 1 yields the following (unnormalized) ensemble of pure states
\begin{align} \label{ensemble1}
    &(1, 1, -1, -1, 1, 1, -1, -1), \notag \\ 
    &(1, -1, 1, -1, -1, 1, -1, 1), \notag \\
    &(1, -1, -1, 1, -1, 1, 1, -1), \notag  \\
    &(1, 1, 1, 1, 1, 1, 1, 1), \notag  \\
    &(0, \tfrac{1}{10}, 0, -\tfrac{7}{10}, -\tfrac{1}{10}, 0, \tfrac{7}{10}, 0), \notag \\
    &(\tfrac{1}{10}, 0, -\tfrac{7}{10}, 0, 0, -\tfrac{1}{10}, 0, \tfrac{7}{10}), \notag \\
    &(-\tfrac{7}{10}, 0, -\tfrac{1}{10}, 0, 0, \tfrac{7}{10}, 0, \tfrac{1}{10}), \notag \\
    &(0, -\tfrac{7}{10}, 0, -\tfrac{1}{10}, \tfrac{7}{10}, 0, \tfrac{1}{10}, 0), 
\end{align}
with respective probabilities $(27, 22, 5, 2, 14, 14, 8, 8)/100$.

For Process 2, the initial state consists of the following ensemble: 
\begin{align}\label{ensemble2}
    &(e^{-2i\pi/3}, e^{-5i\pi/6}, 0, 0, 0, 0, 0, 0, 0, 1, 0, 0)/\sqrt{3}, \notag \\
    &(0, 0, e^{-2i\pi/3}, e^{i\pi/6}, 0, 0, 0, 1, 0, 0, 0, 0)/\sqrt{3}, \notag \\
    &(1, i, 0, 0, 0, 0, 0, 0, \sqrt{3}, 1, 0, 0)/\sqrt{6}, \notag \\ &(0, 0, -1, i, 0, 0, \sqrt{3}, -1, 0, 0, 0, 0)/\sqrt{6}, \notag\\ &(0, 0, 0, 0, 1, 0, 0, 0, 0, 0, 0, 0),
\end{align}
with respective probabilities $(1, 1, 1, 1, 4)/8$. For this ensemble, the system that is sent to Bob is a qutrit. For the first four states above, Bob's qutrit state is restricted to the subspace of the first two levels; the fifth state above consists of a tensor product state $|00\rangle_{AC}$ in product with the third level basis state of Bob's qutrit. 

Using the state preparation module described in the previous section, we prepare the above ensembles for the two processes accordingly.

\subsection{Implementing the POVMs}
In Refs.~\cite{Kurzynski2013,Zhao2015}, a scheme to implement arbitrary POVMs via a discrete-time quantum walk has been proposed. Quantum walks model an evolution of a particle in a discrete space, known as the position state, depending on that of a qubit state, known as the coin state. When the coin state is measured, the interaction between it and the position state leads to the position state moving back and forth. The movement leads to an interference and this interference pattern corresponds to the implementation of a certain POVM on the coin state.

The inset quantum network of Fig.~2 of the main text is used to construct the POVMs $\Theta_B$ and $\Pi_B$. The position state and the coin state are encoded in the longitudinal spatial modes and polarization of the second photon, respectively. The evolution of any step of walk is unitary and consists of a coin toss operation acting on the coin state and a conditional translation operation acting on the position state. The coin toss is realized by an assembly of a HWP and a QWP, and can be represented as a $2\times2$ unitary matrix. The conditional translation operation which the coin controls is realized by a beam displacer and reads
\begin{equation}
T=\sum_x |x+1, V\rangle\langle x, V| + |x-1, H\rangle\langle x, H|.
\end{equation}

The general procedure to construct a POVM with $n$ elements was proposed in Ref.~\cite{Kurzynski2013}: 
\begin{enumerate}
    \item Initiate the quantum walk at position $x=0$ with the coin state corresponding to the qubit state one wants to measure. 
    \item Set i=1. 
    \item While $i<n$, do:
        \begin{enumerate}[label=(\alph*)]
            \item Apply coin operation $C_i^{(1)}$ at position $x=0$ and identity elsewhere and then apply translation operator T. 
            \item Apply coin operation $C_i^{(2)}$ at position $x=1$, bit flip operation at position $x=-1$ and identity elsewhere and then apply translation operator T. 
            \item $i \mapsto i+1$.
        \end{enumerate}
\end{enumerate}

By adjusting the wave plates in the optical network properly, all input photon pass through 4 outputs ports after a 4-step walk process, each of which corresponds to the event of a single element of the POVMs. The coin operators for the POVM $\Theta_B$ are
\begin{align}
C_1^{(1)}&=C_2^{(2)}=C_3^{(1)}=C_3^{(2)}={
\left[ \begin{array}{cc}
1&0\\
0&1\\
\end{array}
\right ]}, \notag \\ 
C_1^{(2)}&=\frac{1}{1+\sqrt{2}}{
\left[ \begin{array}{cc}
\sqrt{2}&1\\
1&-\sqrt{2}\\
\end{array}
\right ]},\notag\\
C_2^{(1)}&={
\left[ \begin{array}{cc}
-2^{\frac{1}{4}}&\sqrt{\frac{\sqrt{2}-1}{\sqrt{2}}}\\
\sqrt{\frac{\sqrt{2}-1}{\sqrt{2}}}&2^{\frac{1}{4}}\\
\end{array}
\right ]}. 
\end{align}

The coin operators for the tetrahedral POVM $\Pi_B$ are
\begin{align}
C_1^{(1)}&=\frac{1}{\sqrt{6+2\sqrt{3}}}{
\left[ \begin{array}{cc}
1+\sqrt{3}&\sqrt{2}\\
\sqrt{2}e^{i\pi/4}&-(1+\sqrt{3})e^{i\pi/4}\\
\end{array}
\right ]}, \notag \\ 
C_1^{(2)}&=\frac{1}{\sqrt{2}}{
\left[ \begin{array}{cc}
-1&1\\
1&1\\
\end{array}
\right ]},\notag\\
C_2^{(1)}&=\frac{1}{\sqrt{2}}{
\left[ \begin{array}{cc}
1&1\\
1&-1\\
\end{array}
\right ]}, \notag \\
C_2^{(2)}&=\frac{1}{\sqrt{3}}{
\left[ \begin{array}{cc}
\sqrt{2}&1\\
1&-\sqrt{2}\\
\end{array}
\right ]}, \notag \\
C_3^{(1)}&=\frac{1}{\sqrt{2}}{
\left[ \begin{array}{cc}
e^{-i\pi/3}&e^{i\pi/6}\\
e^{i\pi/3}&e^{-i\pi/6}\\
\end{array}
\right ]}, \notag \\
C_3^{(2)}&={
\left[ \begin{array}{cc}
1&0\\
0&1\\
\end{array}
\right ]}. 
\end{align}

\section{Experiment Results}\label{app:experiment}

\subsection{Tomography} 
\begin{figure}[b!]
    \centering
    \includegraphics[width=0.9\linewidth]{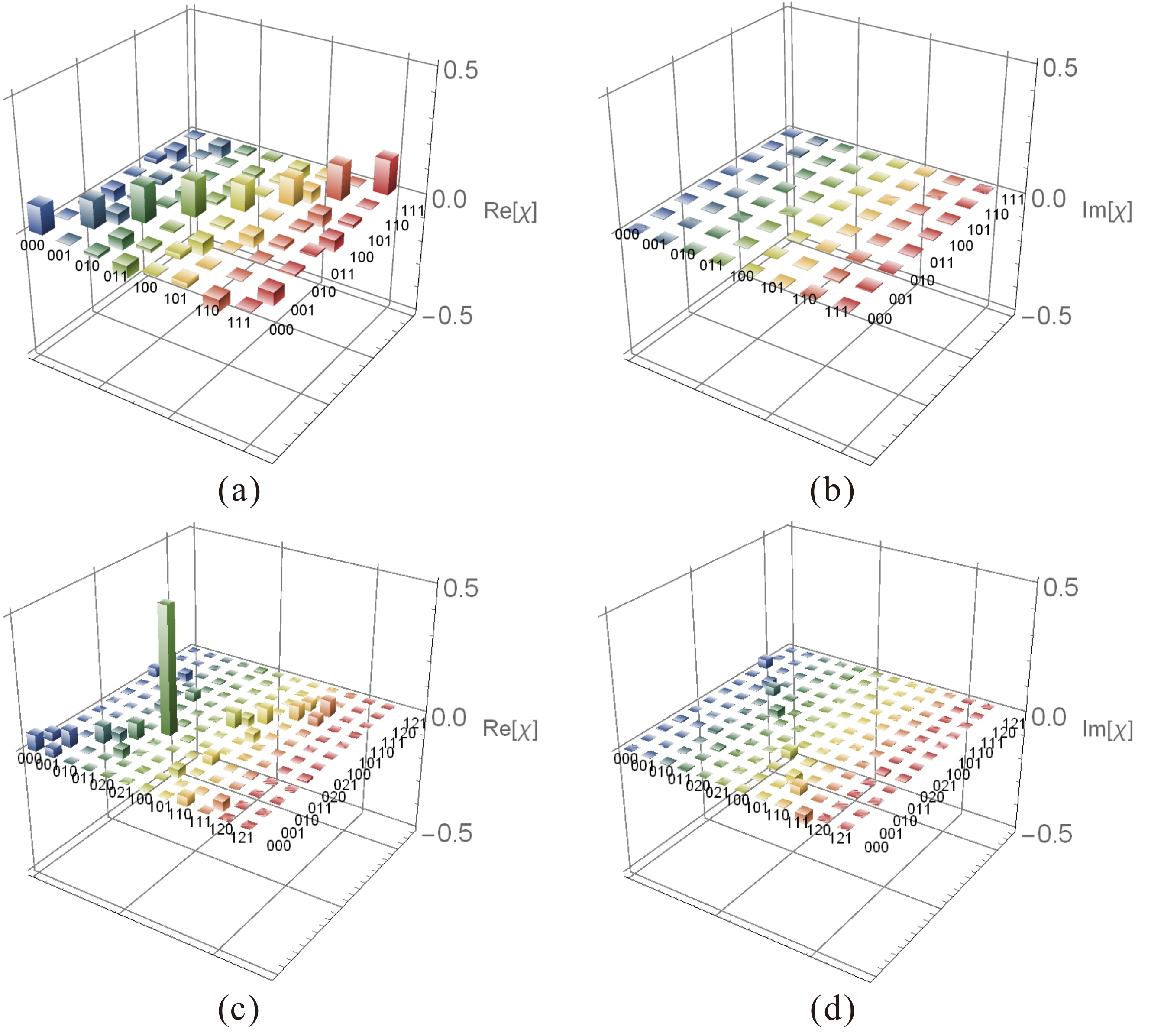}
    \caption{\textit{Tomography results.} (a) Real part and (b) imaginary part of reconstructed density matrix of the state $\lambda_{ABC}$. (c) Real part and (d) imaginary part of reconstructed density matrix of the state $\omega_{ABC}$.}
    \label{fig:tomog}
\end{figure}

We firstly characterize the process tensors $\Lambda_{ABC}$ and $\Omega_{ABC}$ by performing quantum tomography procedure on corresponding quantum states $\lambda_{ABC}$ and $\omega_{ABC}$. The reconstructed density matrices are shown in Fig.~\ref{fig:tomog} and their fidelities $\mathcal{F}(\rho, \sigma) := \left(\tr{\left[\sqrt{\sqrt{\rho} \sigma \sqrt{\rho}}\right]}\right)^2$ to the ideal cases are calculated to be $0.9862\pm0.0005$
and $0.9858\pm0.0008$. The high fidelities show that our technique can effectively construct the desired process tensors which can be obtained through tensor product of these states with identity matrices on the output spaces of the history and memory systems. Based on the reconstructed process tensors, we calculate their non-Markovianity; the results are presented in the main text.
 
\subsection{Markov Order and Memory Strength} 
To demonstrate the memory length of both processes $\Lambda_{ABC}$ and $\Omega_{ABC}$, we analyze the Markov order by measuring the memory strength with respect to different instruments. The instrument-specific memory strength is defined as any suitable correlation monotone between the history and future of the process when the memory is probed by certain instrument. Here, we choose the correlation measure to be the quantum mutual information $I(A:C):=S_A+S_C-S_{AC}$. For Process 1, $\Lambda_{ABC}$, when performing the POVM $\Theta_B$ defined in Eq.~\eqref{eq:appperespovm} on Bob's qubit, the mutual information between Alice and Charlie for each event is $0.004, 0.005, 0.010$ respectively, with a typical error bar of $0.001$, while the theoretical expectation is approximately $0$. Aggregating these to the instrument level (with a uniform average) gives a memory strength of $0.041\pm0.0015$ for the overall instrument. For Process 2, $\Omega_{ABC}$, when performing the tetrahedral POVM on the first two levels of Bob's qutrit [see Eq.~\eqref{eq:appqutritsharp}], the memory strength for each event is $0.216, 0.171, 0.165, 0.188$ respectively, with a typical error bar of $0.010$, while the theoretical expectation is $0.2075$. Aggregating these to the instrument level (with a uniform average) gives a memory strength of $0.185\pm0.010$ for the overall instrument. When choosing the noisy measurement defined in Eq.~\eqref{eq:appfuzzy}, the memory strength for each event is $0.004$ with a typical error bar of $0.002$, while the theoretical expectation is $0$. We also demonstrate the non-vanishing quantum CMI feature when the processes exhibit finite Markov order. For the processes $\Lambda_{ABC}$ and $\Omega_{ABC}$, we observe values of $0.0524\pm0.0014$ and $0.443\pm0.004$, while the theoretical predictions are $0.059$ and $0.5$ respectively.

\begin{figure}[t!]
    \centering
    \includegraphics[width=0.9\linewidth]{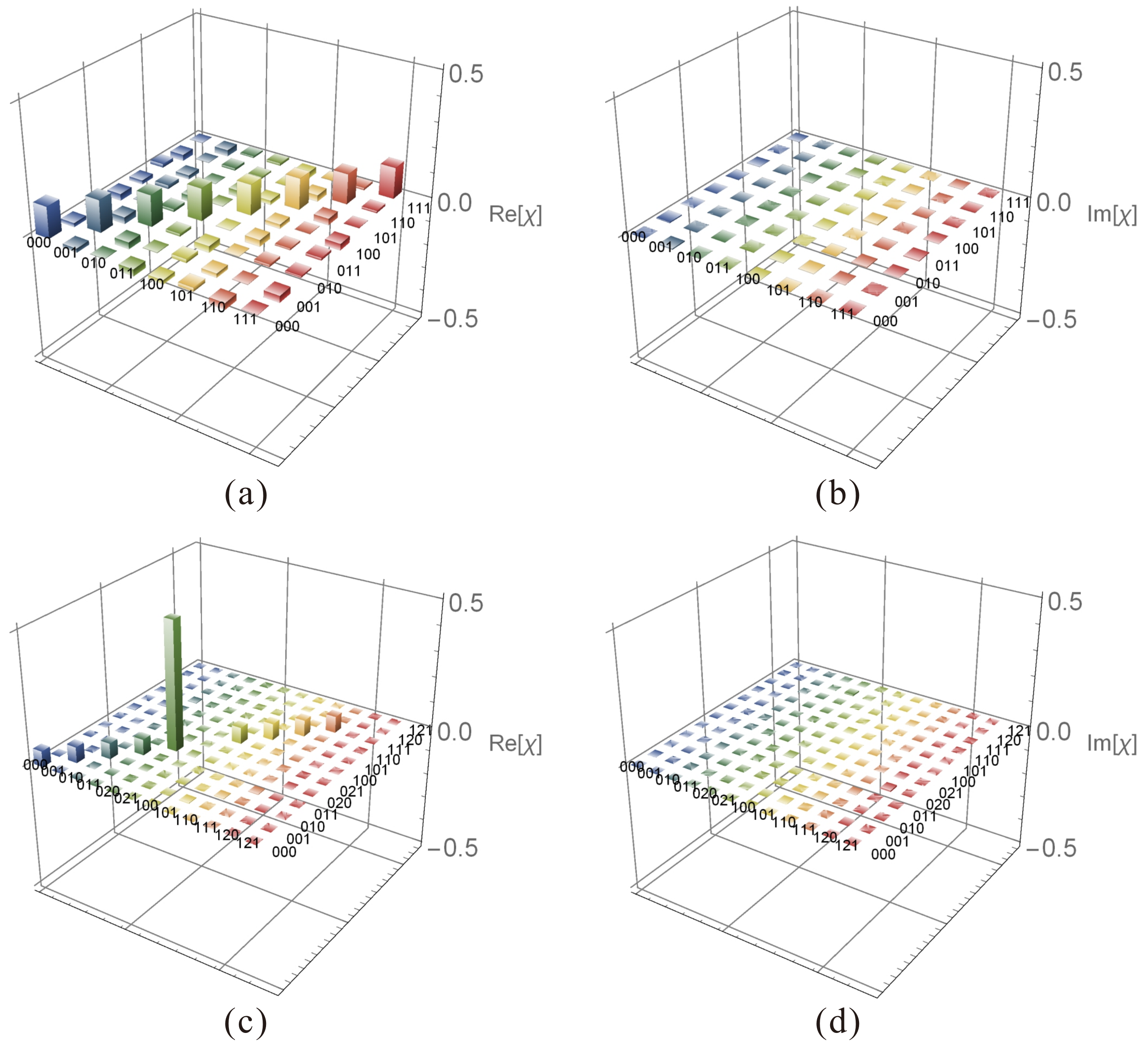}
    \caption{\textit{Approximate reconstruction results.} (a) Real part and (b) imaginary part of reconstructed density matrix of the approximate recovered state $\underline{\lambda}_{ABC}^{\Theta_B}$. (c) Real part and (d) imaginary part of reconstructed density matrix of the approximate recovered state $\underline{\omega}_{ABC}^{\Xi_B}$. These recovered states have no future-history correlations and approximates the true ones $\lambda_{ABC}$ and $\omega_{ABC}$ where the (negligible) future-history correlations are present [see Fig.~\ref{fig:tomog} (a-d)]. The fact that the recovered process here are real diagonal matrices arise from the fact that the history-blocking instruments are POVMs rather than more general instruments that feed-forward the post-measurement state, as will be the case for all common-cause processes with finite-length memory. In the more general case, where the post-measurement states play a role, the recovered processes could display coherences corresponding to temporal quantum effects.}
    \label{fig:simutomog}
\end{figure}
 
\subsection{Efficient Recovery} 

We now detail the procedure for recovering an approximate description for the process with respect to a given history-blocking instrument. This protocol allows one to accurately approximate expectation values of any observable of the form $C_{ABC} = \sum_{y} c_y \mathsf{X}^{(y)}_{ABC}$ where $\mathsf{X}^{(y)} = \sum_{x} \! \mathsf{E}^{(x,y)}_{AC} \!\otimes \mathsf{O}^{(x)}_{B}$, with arbitrary $\mathsf{E}^{(x,y)}_{AC}$, ensuring that Bob's part of the observable lies within the span of said history-blocking instrument~\cite{Taranto2019S}. Intuitively, the approximate process simulates the behavior of the true one for any experiment where Bob implements any linear combination of the history-blocking instruments; conversely, it cannot be used to sensibly predict statistics for general instruments with elements lying outside of this span, as the reconstruction makes use of no such information. Mathematically, such objects are \emph{restricted} process tensors in the sense of Ref.~\cite{Milz2018A}; this means that they are not positive operators in general, but nonetheless act as positive operators on the subspace spanned by the instrument. We notationally denote this in the following by underlining the recovered objects to remind the reader that care must be taken regarding the scenarios in which they can be sensibly used. We now highlight the accuracy of this approximation by experimentally reconstructing the recovered process for both processes considered. 

As the memory strength of Process 1, $\Lambda_{ABC}$, with respect to the POVM $\Theta_B$ in Eq.~\eqref{eq:appperespovm} is small, we can reconstruct the recovered process $\underline{\Lambda}_{ABC}^{\Theta_B}$ from our experimental data. As per Ref.~\cite{Taranto2019S}, the recovered process is a compressed description that discards the negligible future-history correlations: we take the conditional Alice-Charlie states for each of the measurements events, $\{ \lambda_{AC}^{(x)} \}$, and replace them by the tensor product of their marginals, $\{ \lambda_{A}^{(x)} \otimes \lambda_{C}^{(x)} \}$. We then construct the dual set of operators for Bob's measurement $\Theta_B$; these are: 
\begin{align}
    \Delta_B^{(1)} &= \frac{1}{2}(\ket{0}+\ket{1})(\bra{0}+\bra{1})-\frac{1+\sqrt{2}}{2}(\ket{0}\bra{0}-\ket{1}\bra{1}) \notag \\
    &= \frac{1}{2} (-\sqrt{2}\ket{0}\bra{0} + \ket{0}\bra{1} + \ket{1}\bra{0} + (2+\sqrt{2})\ket{1}\bra{1})\notag \\
    \Delta_B^{(2)} &=\frac{1+\sqrt{2}}{2}(\ket{0}-\ket{1})(\bra{0}-\bra{1})+\frac{1-\sqrt{2}}{2}\ket{0}\bra{0} \notag \\
    &-\frac{1+\sqrt{2}}{2}\ket{1}\bra{1} \notag \\
    &= \ket{0}\bra{0} - \frac{1}{2} (1+\sqrt{2}) ( \ket{0}\bra{1} + \ket{1}\bra{0}) \notag \\
    \Delta_B^{(3)} &=\frac{1}{2}[(\ket{0}+\ket{1})(\bra{0}+\bra{1})+\ket{0}\bra{0}-\ket{1}\bra{1}] \notag \\
    &= \ket{0}\bra{0} + \frac{1}{2}(\ket{0}\bra{1} + \ket{1}\bra{0}).
\end{align} 
These operators satisfy $\tr{ [\Delta_B^{(x)} \Theta_B^{(y) \textup{T}}]} = \delta_{xy}$, ensuring the correct statistics for Bob's measurement outcomes for said instrument, and, by linearity, for all linear combinations thereof. By taking the tensor product of Alice and Charlie's local marginal states for each measurement outcome with the dual elements to Bob's instrument on all of the input spaces, weighted by the appropriate probabilities, we yield the approximate recovered state $\underline{\lambda}_{ABC}^{\Theta_B}$ (see Fig.~\ref{fig:simutomog} (a-b) for the tomography results) which leads to the recovered process $\underline{\Lambda}_{ABC}^{\Theta_B}=\underline{\lambda}_{A^\inp B^\inp C^\inp}^{\Theta_B}\otimes \mathbbm{1}_{A^\out B^\out}$. The fidelity of the recovered state constructed from our data to ideal case
\begin{align}
   \underline{\lambda}_{A^\inp B^\inp C^\inp}^{\Theta_B}&= 2(3-2\sqrt{2})\left[\lambda_{A^i}^{(1)}\otimes\Delta_{B^\inp}^{(1)}\otimes\lambda_{C^i}^{(1)}\right. \notag \\ &+\left.\lambda_{A^i}^{(2)}\otimes\Delta_{B^\inp}^{(2)}\otimes\lambda_{C^i}^{(2)}\right] \notag \\ &+(8\sqrt{2}-11)\lambda_{A^i}^{(3)}\otimes\Delta_{B^\inp}^{(3)}\otimes\lambda_{C^i}^{(3)},
\end{align}
with $\{ \lambda_X^{(x)} \}$ in Eq.~\eqref{eq:applambda}, is $0.9979\pm0.0014$.
 
As the memory strength of Process 2, $\Omega_{ABC}$, for the noisy measurement $\Xi_B$ is small, we can similarly reconstruct a recovered process $\underline{\Omega}^{\Xi_B}_{ABC}$ from our experimental data that accurately approximates the actual one. As per Ref.~\cite{Taranto2019S}, we take the conditional Alice-Charlie states for each of the two noisy measurements events, $\{ \omega_{AC}^{(1)}, \omega_{AC}^{(2)} \}$ and replace them by the tensor product of their marginals, $\{ \omega_{A}^{(1)} \otimes \omega_{C}^{(1)}, \omega_{A}^{(2)} \otimes \omega_{C}^{(2)} \}$. For the process considered, this should yield $\{ \tfrac{\mathbbm{1}_A}{2} \otimes \tfrac{\mathbbm{1}_C}{2}, \ket{0}\bra{0}_{A} \otimes \ket{0}\bra{0}_C \}$. Since the events of the noisy measurement that Bob makes correspond to $\{ \mathbbm{1}^{01}, \ket{2}\bra{2}\}$ (where the superscript labels the subspace of Bob's qutrit on which the identity acts), which is a self-dual set of operators (up to normalization by dimension $d$ of the subspace on which the operators act) such that $\tr{[\Xi_B^{(x) } \Xi_B^{(y) \textup{T}}]} = \delta_{xy} d$, we can reconstruct the initial state relevant for the recovered process from our data by taking the tensor product of the conditional marginal states on Alice and Charlie's input spaces $\{ \omega_{A^\inp}^{(1)} \otimes \omega_{C^\inp}^{(1)}, \omega_{A^\inp}^{(2)} \otimes \omega_{C^\inp}^{(2)} \}$ with the suitably normalized elements of Bob's measurement on the input space, yielding the recovered state (see Fig.~\ref{fig:simutomog} (c-d) for the tomography results)
\begin{align}\label{eq:recoveredstate}
    \underline{\omega}_{A^\inp B^\inp C^\inp}^{\Xi_B} = \frac{1}{2} \left( \omega_{A^\inp}^{(1)} \otimes \mathbbm{1}^{01}_{B^\inp} \otimes \omega_{C^\inp}^{(1)} + \omega_{A^\inp}^{(2)} \otimes \ket{2}\bra{2}_{B^\inp} \otimes \omega_{C^\inp}^{(2)} \right).
\end{align}
This state is (by construction) steered into an uncorrelated Alice-Charlie state for each of Bob's measurement events. Taking the tensor product of this with identity operators on Alice and Bob's output spaces yields the recovered process $\underline{\Omega}_{ABC}^{\Xi_B}=\underline{\omega}_{A^\inp B^\inp C^\inp}^{\Xi_B}\otimes \mathbbm{1}_{A^\out B^\out}$. The fidelity of the recovered state constructed from our data in Eq.~\eqref{eq:recoveredstate} to ideal case
\begin{align}
    \frac{1}{2} \left( \frac{\mathbbm{1}_{A^\inp}}{2} \otimes \mathbbm{1}^{01}_{B^\inp} \otimes \frac{\mathbbm{1}_{C^\inp}}{2} + \ket{0}\bra{0}_{A^\inp} \otimes \ket{2}\bra{2}_{B^\inp} \otimes \ket{0}\bra{0}_{C^\inp} \right)
\end{align}
is $0.9960\pm0.0011$. 

To further show the efficiency of the recovered process, we approximate the expectation value of a class of multi-time observables on it and compare the result with the one on the actual process. Here we calculate the expectation value difference of observables on the recovered process and the true process where Alice and Charlie are restricted to perform qubit projective measurements and Bob performs any non-selective measurement; as this operation lies in the span of any measurement for Bob, it is in particular valid to simulate expectation values for observables of this form for both approximate processes that we have reconstructed. This scenario can be reduced to four real parameters $(\theta_1, \phi, \theta_2, \psi)$ by parameterizing Alice's and Charlie's projectors as $\{|\phi\rangle\langle\phi|, |\phi\rangle\langle\phi|^\perp, |\psi\rangle\langle\psi|,|\psi\rangle\langle\psi|^\perp\}$, where $|\phi\rangle=\cos{\theta_1}|0\rangle+e^{i \phi}\sin{\theta_1}|1\rangle$, $|\phi\rangle^\perp=\sin{\theta_1}|0\rangle-e^{-i \phi}\cos{\theta_1}|1\rangle$ and $|\psi\rangle=\cos{\theta_2}|0\rangle+e^{i \psi}\sin{\theta_2}|1\rangle$, $|\psi\rangle^\perp=\sin{\theta_2}|0\rangle - e^{-i \psi}\cos{\theta_2}|1\rangle$. For Process 1, the maximum value of the difference is $0.048$, obtained with $\theta_1=0.250\pi$, $\theta_2=0.190\pi$, $\phi=0$, and $\psi=0$. 
For Process 2, the maximum value of the difference is $0.022$, obtained with $\theta_1=0.120\pi$, $\theta_2=0.200\pi$, $\phi=1.920\pi$, and $\psi=\pi$. See Fig.~4 in the main text.

\end{document}